# Resonant Electric-Magnetic Toroidal Duality in Height-Modulated Hexagonal Metasurfaces

**Oleksiy Breslavets[1*], Yuri Savin[1] and Zoya Eremenko[1,2]**

[1] Department of Acoustic and Electromagnetic Spectroscopy, O. Ya. Usikov Institute for Radiophysics and Electronics of the National Academy of Sciences of Ukraine, 61085 Kharkiv, Ukraine
[2] Department of Solid-State Research, Leibniz Institute for Solid State and Materials Research, Helmholtzstraße 20, 01069 Dresden, Germany

*Author to whom any correspondence should be addressed.

**E-mail:** alex.breslavets@gmail.com

**Keywords:** toroidal modes, all-dielectric metasurfaces, hexagonal lattice, quasi-BICs, high-Q resonances, terahertz photonics

## Abstract

Objective: Toroidal modes enable high-Q resonances, but electric toroidal excitations remain unexplored compared to magnetic ones. This work establishes electric-magnetic toroidal duality in a hexagonal metasurface.
Approach: Using finite-element simulations, we analyze electric and magnetic toroidal modes in a hexagonal silicon nanorod supercell under mirror-symmetry breaking via height modulation. Eigenfrequencies, Q-factors, power flow, and polarization responses are computed.
Main results: We identify electric TO and ATO modes with complementary near-field topologies to magnetic analogues. Direct frequency intersections (magnetic and electric TO/ATO) yield high-Q quasi-BICs. Polarization selectivity reverses between families: 0° excites magnetic TO/electric ATO; 90° excites magnetic ATO/electric TO. A loss hierarchy (magnetic TO > magnetic ATO > electric ATO > electric TO) and protective layers compatibility are demonstrated.
Significance: Electric and magnetic toroidal responses are dual manifestations of the same symmetry, providing a unified design framework for high-Q metasurfaces in sensing, nonlinear optics, and loss-tolerant devices.

## 1. Introduction

All-dielectric metasurfaces (ADMs) support resonant modes with strong field confinement and intrinsically low dissipative losses, enabling access to high quality (Q-) factors essential for applications in nonlinear optics, sensing, and quantum photonics [1–16]. Among the variety of resonant excitations in ADMs, toroidal modes occupy a special position: originating from poloidal displacement currents that form closed field loops, they exhibit weak coupling to the radiation continuum, suppressed far-field scattering, and enhanced near-field localization. Collective arrangements such as trimer clusters further enrich the multipolar spectrum, supporting both toroidal (TO) and antitoroidal (ATO) configurations characterized by uniform or alternating circulation of magnetic field loops, respectively [17–19].

However, this understanding has been overwhelmingly focused on magnetic toroidal responses [19–27], leaving their electric counterparts, where the toroidal moment arises from circulating electric displacement currents forming closed electric field loops, predicted by symmetry but largely unexplored. A systematic comparative analysis of electric and magnetic toroidal excitations within the same structural platform has not been performed to date.

A key challenge lies in the symmetry-protected nature of toroidal states: in perfectly symmetric metasurfaces, these modes are strictly dark, requiring controlled symmetry breaking to unlock radiative coupling while preserving high Q-factors [1]. Such strategies are closely tied to quasi-bound states in the continuum (quasi-BICs), which enable extreme field confinement [4]. Previous works have primarily addressed isolated magnetic toroidal resonances or hybridization with dipolar modes. The possibility of configuring direct frequency intersections between electric and magnetic toroidal modes within the same metasurface (and comparing their behaviors under identical symmetry-breaking conditions) has not been systematically investigated, despite its potential for dual-polarization high-Q platforms.

In this work, we provide a systematic comparative analysis of electric and magnetic toroidal eigenmodes in an all-dielectric metasurface with a hexagonal supercell geometry. We identify and classify both mode types and investigate their dependence on controlled height-induced symmetry breaking. Rather than relying on complex in-plane patterning, we exploit a single out-of-plane symmetry-breaking parameter (nanorod height modulation) to independently tune electric and magnetic toroidal resonances, simplifying fabrication while enabling control over mode hybridization, Q-factors, and polarization selectivity. We demonstrate direct frequency intersections between electric TO and ATO modes, yielding high-Q quasi-BICs, and establish a resonant electric-magnetic toroidal duality: two sides of the same symmetry coin.

## 2. Methods

The electromagnetic response of the metasurface was simulated using the finite-element method (COMSOL Multiphysics). The unit cell model parameters are described below. The supercell period is chosen to ensure subwavelength operation, suppressing higher-order diffraction channels and enabling strong field confinement [23,24]. Operation near 1550 nm is advantageous for sensing applications, as resonance shifts are highly sensitive to refractive index variations introduced by functional coatings [21]. With this design rationale, the hexagonal supercell period is fixed at $p$ = 1148 nm, containing six identical trimer clusters arranged with 60° rotational symmetry (figure 1). Each trimer is composed of three cylindrical silicon nanorods with permittivity $\varepsilon_d$ = 12 - $i$0.001 [28], radius $r_d$ = 170 nm, and initial height $h_d$ = 149 nm. The trimers are rotated by 30° with respect to the $x$-axis. Symmetry breaking is introduced by modifying the height of one nanorod in each trimer by $\Delta h_d/h_d$, implementing a $C_s^{(1)}$ mirror-symmetry perturbation scheme. The structure is considered free-standing/supported by substrate (Si or $SiO_2$ with $\varepsilon_{sub}$ = 2.1 - $i1\times10^{-8}$ [29]).

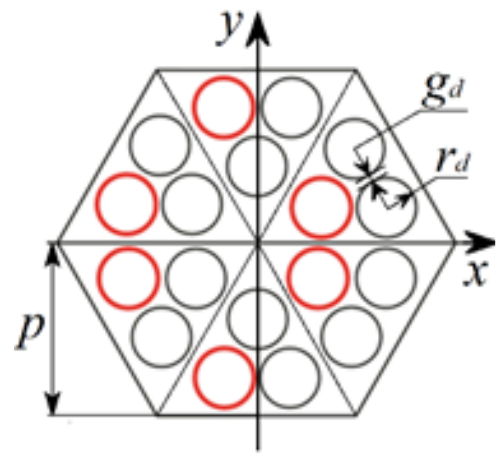

**Figure 1.** Symmetry-breaking scheme [25]: $C_s^{(1)}$ and for ADM supercell. Red circles indicate nanorods with variable heights $h_d \pm \Delta h_d$. Geometrical parameters of the cell are: $\lambda_0/p$ = 1.07 ($\lambda_0$ – operating wavelength of 1228 nm), $r_d/p$ = 0.15, $h_d/p$ = 0.13, $g_d/p$ = 0.04, $r_t/p$ = 0.67, $h_{sub}/p$ = 0.13.

Numerical simulations were performed using COMSOL Multiphysics. Accordingly, we take the imaginary part of the permittivity with a negative sign, in accordance with the COMSOL notation for the time dependence *exp(iωt)*. Eigenmode calculations used the Eigenfrequency Solver with Floquet periodic boundary conditions in the lateral directions and perfectly matched layers (PMLs) along the out-of-plane axis. Frequency-domain simulations were carried out under normal-incidence plane-wave excitation to compute reflection and transmission coefficients as functions of frequency, asymmetry parameter $\Delta h_d/h_d$, implementing a $C_s^{(1)}$, and azimuth polarization angle $\varphi$. Q-factors were extracted from spectral linewidths via Lorentzian or Fano fitting. All collective eigenmodes are in Supplementary Material, section S8, table 7.

## 3. Results

### 3.1. Eigenmode spectrum of hexagonal metasurfaces

We analyze the eigenmode spectrum of the hexagonal trimer metasurface in the absence of symmetry breaking ($\Delta h_d/h_d$ = 0). The classification of eigenmodes supported by trimer-based dielectric metasurfaces has been extensively developed in [17,20,25–27]. The trimer-level behavior has been discussed in [17,25,26]. Among the collective excitations supported by the structure, particular attention is paid to TO and ATO modes due to their weak radiative coupling and strong near-field localization. Representative magnetic and electric field distributions of the magnetic TO and ATO eigenmodes [18,19,21,23–26] are shown in figure 2, and electric TO and ATO modes are presented in figure 3. In the magnetic TO mode, the magnetic field forms uniform closed in-plane circulation loops across all trimers, while in the magnetic ATO mode the circulation alternates between neighboring trimers (figure 2). In contrast, the electric TO mode is characterized by closed electric field loops localized within the inter-nanorod gaps of each trimer, forming a triangular toroidal pattern, while the electric ATO mode exhibits alternating circulation of the electric field (figure 3). These field topologies confirm the toroidal nature and dark-state character of all four modes. The transition from an isolated trimer or two-dimensional configuration to the three-dimensional hexagonal supercell significantly increases the Q-factors (The Q-factors are not explicitly quantified in [17,20,25–27]) with values reaching $10^6$-$10^8$ compared to $10^2$-$10^3$ in lower-dimensional structures (see Supplementary Material, section S8, tables 1-6,9). This improvement reflects symmetry-enabled suppression of radiation losses. At the symmetric point ($\Delta h_d/h_d$ = 0), the magnetic TO and ATO modes exhibit eigenfrequencies of 255.5 THz and 242.73 THz, respectively, with Q-factors of $9.8\times10^7$ and $1.3\times10^7$.

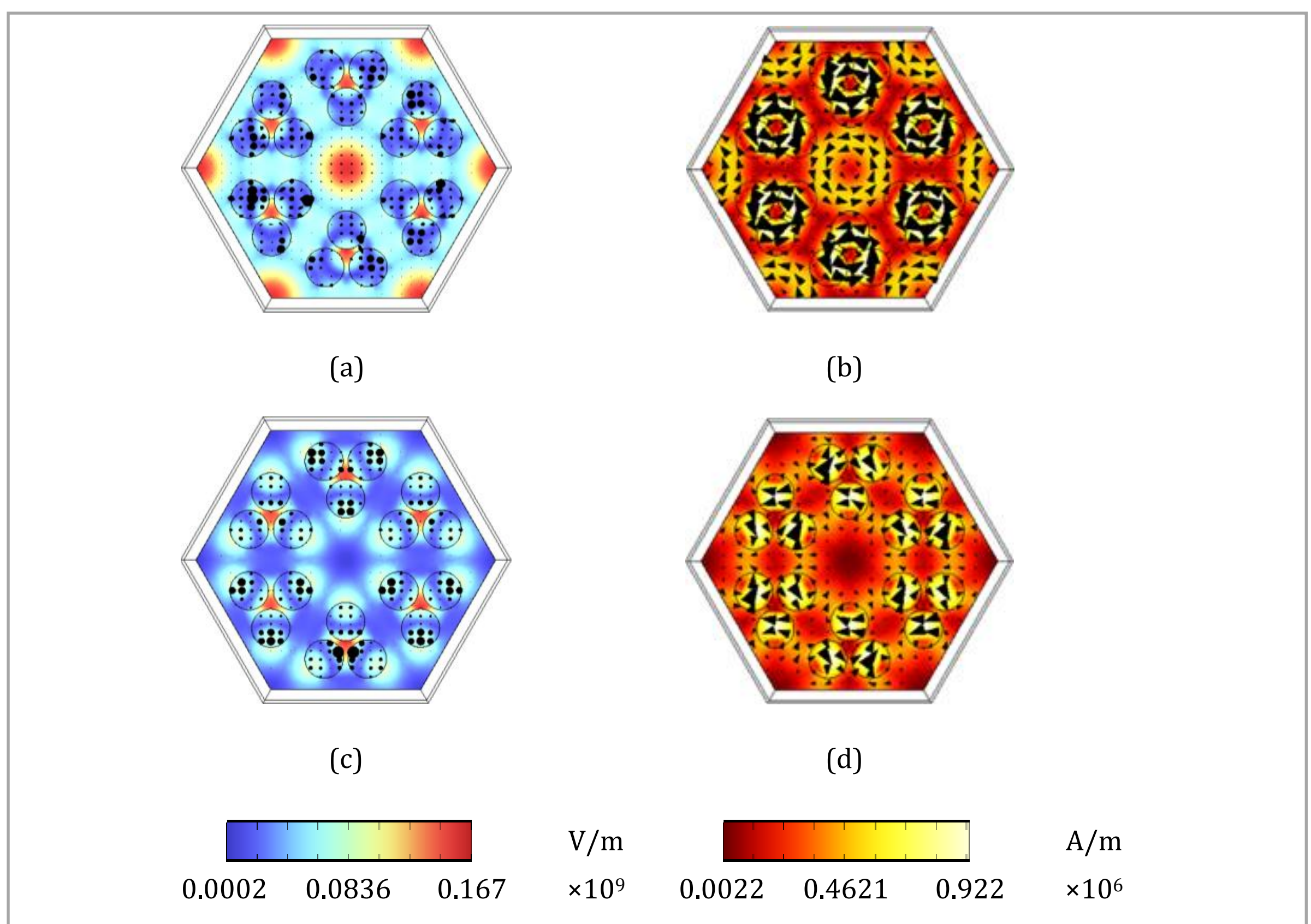

**Figure 2.** Eigenmode (a),(b) and (c),(d) electric and magnetic fields of the magnetic TO and ATO modes, respectively, in 3D hexagonal supercell with initial $h_d$.

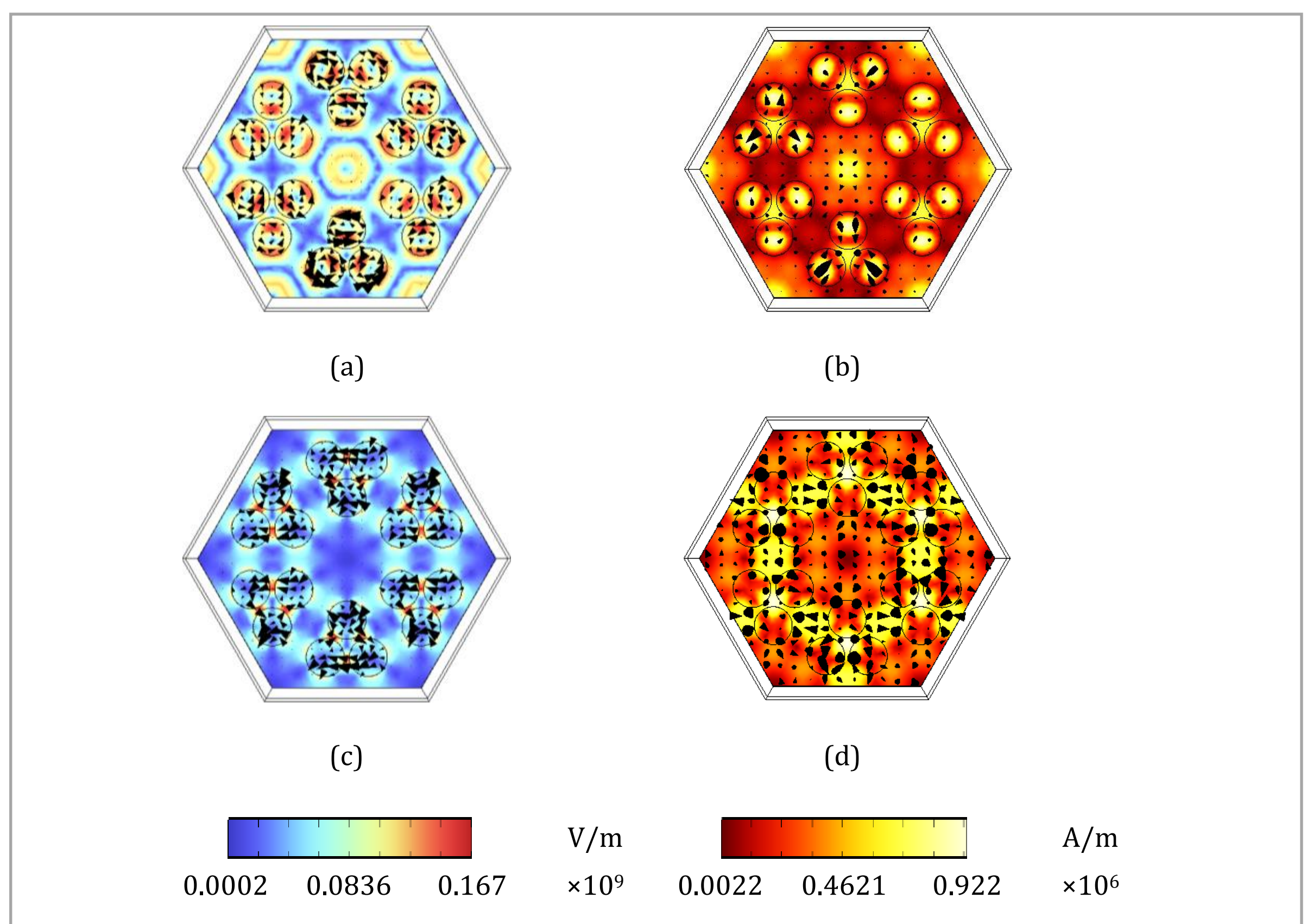

**Figure 3.** Eigenmode (a),(b) and (c),(d) electric and magnetic fields of the electric TO and ATO modes, respectively, in 3D hexagonal supercell with initial $h_d$. Styles and ranges match figure 2.

The electric TO and ATO modes appear at higher frequencies: 243.76 THz ($Q$ = 2.2×10$^8$) and 247.8 THz ($Q$ = 1.3×10$^7$), respectively. Notably, the electric TO mode exhibits a Q-factor approximately one order of magnitude higher than its magnetic counterpart, indicating stronger field confinement and reduced radiation losses. This disparity highlights the higher sensitivity of electric toroidal resonances to symmetry and field localization, enabling superior modal selectivity.

A direct comparison of Figures 2 and 3 reveals systematic similarities between the two mode families. The magnetic field of the magnetic ATO mode and the electric field of the electric TO mode share identical minima at the cell center, trimer corners, and inter-trimer regions. Conversely, the magnetic field of the magnetic TO mode and the electric field of the electric ATO mode both exhibit ring-shaped maxima along the trimer circumferences—a characteristic signature of TO-type modes.

Furthermore, the electric field of the magnetic ATO mirrors the magnetic field of the electric TO (central maxima within trimers, minima between them), while the electric field of the magnetic TO mirrors the magnetic field of the electric ATO (central maxima with strong inter-trimer fields). Across both families, ATO resonances occur at lower frequencies than TO resonances, and the relative Q-factor ordering follows the same paired relationship. These consistent correspondences reinforce the fundamental interconnectedness between electric and magnetic toroidal excitations.

Comparison of the field distributions in figures 2 and 3 reveals a consistent duality: the magnetic field of the magnetic ATO mode mirrors the electric field of the electric TO mode, while the electric field of the magnetic TO mode corresponds to the magnetic field of the electric ATO mode. This complementary near-field topology establishes the fundamental duality between electric and magnetic toroidal excitations.

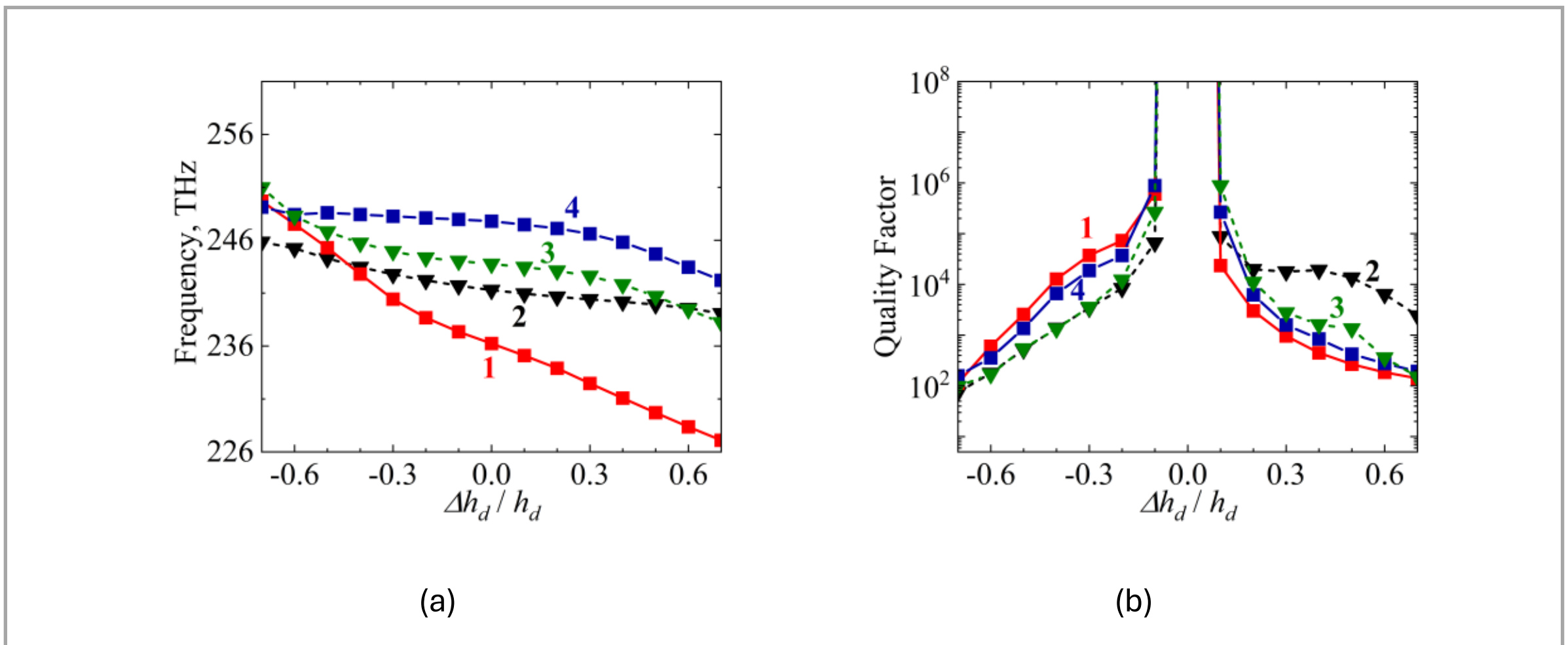

**Figure 4.** Dependence of (a) eigenfrequencies and (b) Q-factors of the four toroidal modes on the relative height asymmetry parameter $\Delta h_d/h_d$ for the free-standing metasurface under $C_s^{(1)}$ symmetry breaking. Lines: (1) magnetic ATO, (2) magnetic TO, (3) electric TO, (4) electric ATO. At $\Delta h_d/h_d$ = 0, all modes are symmetry-protected dark states with extremely high Q-factors. Intersection points are indicated by vertical dashed lines: magnetic ATO/TO at $\Delta h_d/h_d$ = -0.44, electric ATO/TO at $\Delta h_d/h_d$ = -0.62, and magnetic ATO/electric ATO at $\Delta h_d/h_d$ = -0.65.

### 3.2. Symmetry-controlled interaction and intersection of TO and ATO modes

We now examine how the eigenfrequencies and Q-factors of the four toroidal modes evolve with the height-asymmetry parameter $\Delta h/h$ , as shown in figure 4.

At the symmetric point ($\Delta h/h$ = 0), all modes are symmetry-protected dark states with formally infinite Q-factors. Introducing a small asymmetry lifts the dark-state condition, causing the Q-factors to decrease symmetrically as $|\Delta h/h|$ increases, while the eigenfrequencies shift continuously. In both the magnetic and electric modes, ATO resonances occur at lower frequencies than TO resonances, and the relative ordering of their Q-factors follows the same paired relationship, indicating a consistent TO/ATO modal correspondence across electric and magnetic channels.

The frequency evolution reveals three distinct intersection points where two modes become degenerate. The magnetic ATO and TO modes intersect at $\Delta h/h$ = -0.44, the electric ATO and TO modes intersect at $\Delta h/h$ = -0.62, and the magnetic ATO and electric ATO modes intersect at $\Delta h/h$ = -0.65. These accidental degeneracies are of particular interest, as they enable hybridization of distinct toroidal current configurations, giving rise to high-Q quasi-bound states in the continuum (quasi-BICs).

At these intersection points, the electromagnetic field distributions of the intersecting modes exhibit strong spatial overlap, indicating efficient mode hybridization. To illustrate this, Figure 5 presents the power flow distributions (Poynting vector ($P = 0.5 Re(E \times H^*)$[30])) in the *xy*-plane for the magnetic and electric intersection regimes. The energy circulation remains predominantly confined within the trimer clusters, but enhanced leakage into the surrounding medium is observed, consistent with the finite Q-factors. Importantly, the power flow maxima of the TO and ATO modes spatially overlap at the intersection points, confirming the hybridization of the two toroidal configurations at a single frequency.

Recent works provide experimental verification of close geometrically structure with ceramic prototypes confirmed that changing the nanorod height in selected positions of the hexagonal oligomer induces the crossing of TO and ATO resonances. Near-field mapping revealed that ATO modes concentrate energy within trimers, while TO modes localize it in the central cavity region [19].

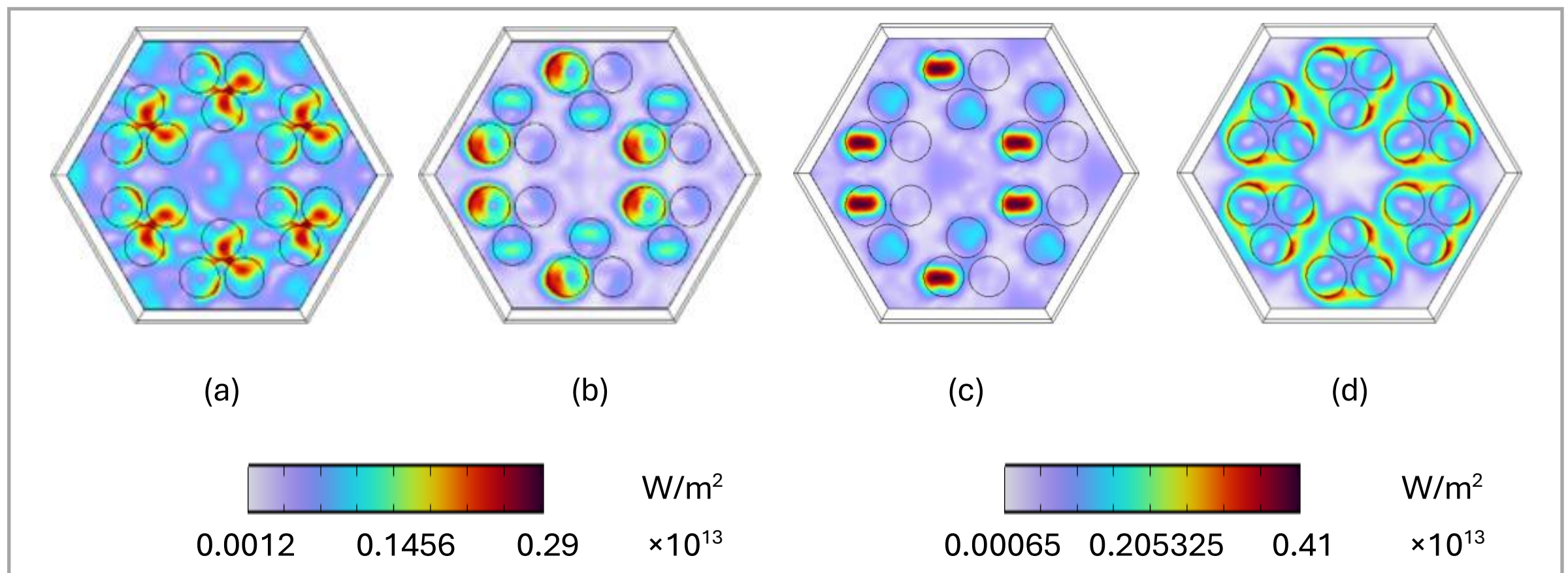

**Figure 5.** Resonant power flow distributions in the *xy*-plane at the metasurface plane for the four toroidal modes at their respective intersection points. (a) Magnetic TO and (b) magnetic ATO modes at the magnetic intersection point $\Delta h/h$ = -0.44. (c) Electric TO and (d) electric ATO modes at the electric intersection point $\Delta h/h$ = -0.62. The power flow patterns reveal spatial overlap of TO and ATO energy circulation at each degeneracy, confirming hybridization of the two toroidal configurations at a single frequency.

At $\Delta h/h$ = 0 the ATO and TO modes are symmetry-protected dark states with purely real eigenfrequencies (formally infinite Q-factors). When a small $\Delta h$ is introduced, both modes acquire finite radiation losses: the Q-factor plot spans from $10^1$ up to $10^6$, indicating extremely high-Q behavior for small perturbations. Specifically, the ATO mode exhibits the highest $Q$ values (reaching towards the $10^6$ scale near minimal asymmetry parameter), whereas the TO mode remains significantly lower, which means roughly one magnitude order smaller across the range ($Q_{ATO}$ / $Q_{TO} \approx 10$) [19].

These eigenmode results display the eigenmode frequency branches $f_{ATO}(\Delta h)$ and $f_{TO}(\Delta h)$ and marks an accidental degeneracy (horizontal dashed line) where $f_{ATO} = f_{TO}$ for a specific $\Delta h$.

The corresponding magnetic near-field patterns in eigenmode results figure illustrate that the ATO mode concentrates magnetic loops and electric near-field maxima inside individual trimers (center regions), while the TO mode concentrates the field predominantly in the central empty cavity of the hexagonal oligomer; arrows in the figure indicate the orientation of toroidal dipole moments in each trimer.

These quantitative ranges are -1.5 mm ≤ $\Delta h$ ≤ +1.5 mm, 8.5–10.5 GHz band, and Q-factor of $10^1$–$10^6$ with an order-of-magnitude Q-factor difference between ATO and TO, that underpin our statements about achievable spectral tuning and the feasibility of polarization-controlled switching and sensing [19]. This single-parameter control stands in contrast to complex multi-parameter optimization schemes typical of toroidal metasurface design.

### 3.3. Azimuth-polarization control of toroidal resonances

After establishing the eigenmode spectrum and symmetry-controlled interaction, we analyze how these modes appear in measurable reflection and transmission characteristics.

Figure 6 presents normalized reflection and transmission responses under normal incidence ($\varphi$ = 0°) for three representative asymmetry values: the symmetric point ($\Delta h_d/h_d$ = 0), the magnetic TO/ATO intersection ($\Delta h/h$ = -0.44), and the electric TO/ATO intersection ($\Delta h_d/h_d$ = -0.62). Increasing symmetry breaking progressively enhances radiative coupling, transforming initially dark states into externally accessible resonances with controllable linewidths – a hallmark of symmetry-enabled quasi-bound states in the continuum [4,5,31–36].

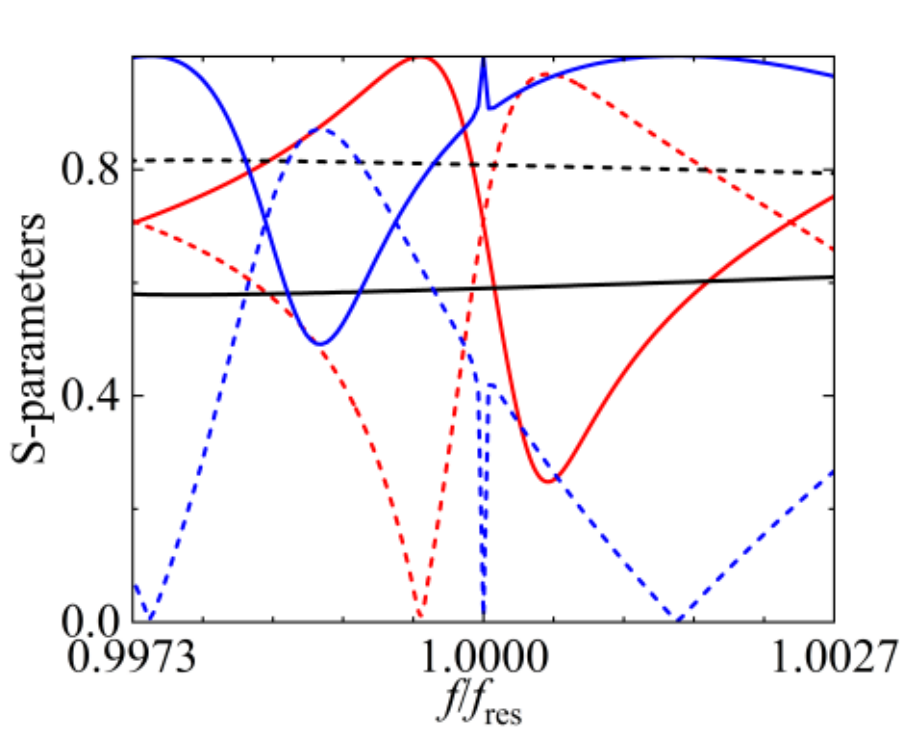

**Figure 6.** Normalized reflection (solid) and transmission (dashed) repones under normal incidence ($\varphi$ = 0°) as a function of $\Delta h/h$ for the $C_s^{(1)}$. The responses are shown for (black line) $\Delta h/h$ = 0 (highest eigen Q-factor state, $f_{res}$= 241.29 THz), (red line) $\Delta h/h$ = -0.44 (magnetic TO/ATO intersection point, $f_{res}$ = 243.79 THz), and (blue line) $\Delta h/h$ = -0.62 (electric TO/ATO intersection point, $f_{res}$ = 244.94 THz for TO and 248.27 THz for ATO).

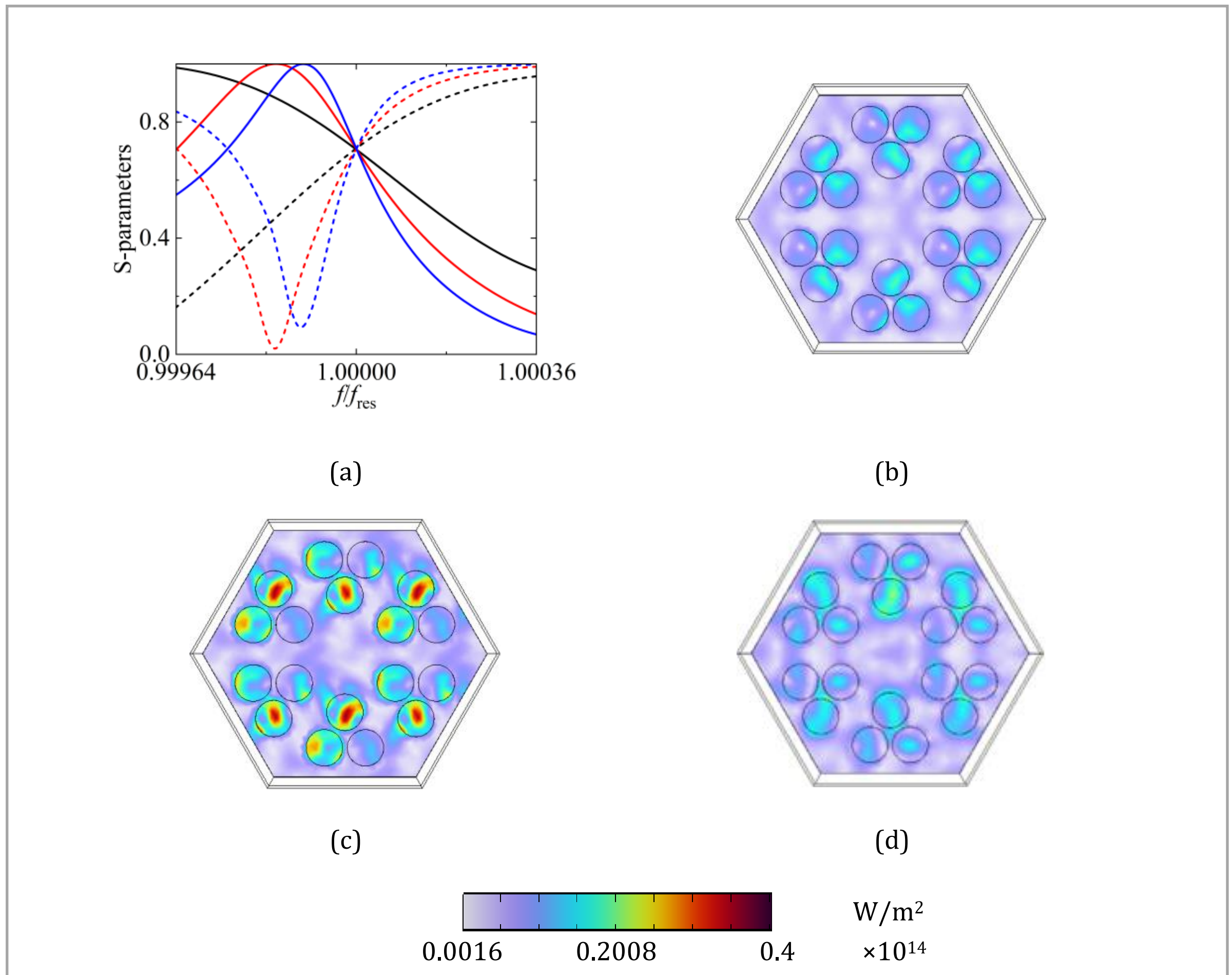

**Figure 7.** Polarization-dependent excitation at the magnetic intersection point ($\Delta h/h$ = -0.44, $f_{res}$ = 243.79 THz). (a) Reflection and transmission responses for azimuth polarization angles $\varphi$ = (black line) 0°, (red line) 45° and (blue line) 90°. The 0° polarization preferentially excites the magnetic TO mode, while 90° favors the magnetic ATO mode. (b)-(d) Resonant power flow distributions in the *xy*-plane for $\varphi$ = 0°, 45° and 90°, respectively.

The excitation process is further controlled by the azimuth polarization angle of the incident wave. Figure 7 focuses on the magnetic intersection regime ($\Delta h/h$ = -0.44). The reflection and transmission responses in figure 7(a) reveal that 0° polarization preferentially excites the magnetic TO mode, while 90° favors the magnetic ATO mode. The Q-factor increases with polarization angle, accompanied by reduced linewidth, indicating enhanced radiative suppression. At 45°, both contributions interfere, producing a mixed response. Corresponding power flow distributions in figure 7(b)-(d) show that the maximum power flow intensity occurs at 45° in figure 7(c), indicating interference of the two orthogonal polarizations.

Figure 8 presents the analogous analysis for the electric intersection regime ($\Delta h_d/h_d$ = -0.62). Here, the polarization selectivity is reversed: 0° polarization preferentially excites the electric ATO mode, while 90° favors the electric TO mode. In contrast to magnetic modes, in figure 8(a), the Q-factor decreases with increasing polarization angle, accompanied by progressive linewidth broadening, revealing opposite polarization-dependent behavior between electric and magnetic toroidal resonances. In contrast to the magnetic case, the power flow distributions reveal that the maximum intensity occurs at 90° in figure 8(d), highlighting the opposite coupling behavior of electric toroidal modes to incident field polarization.

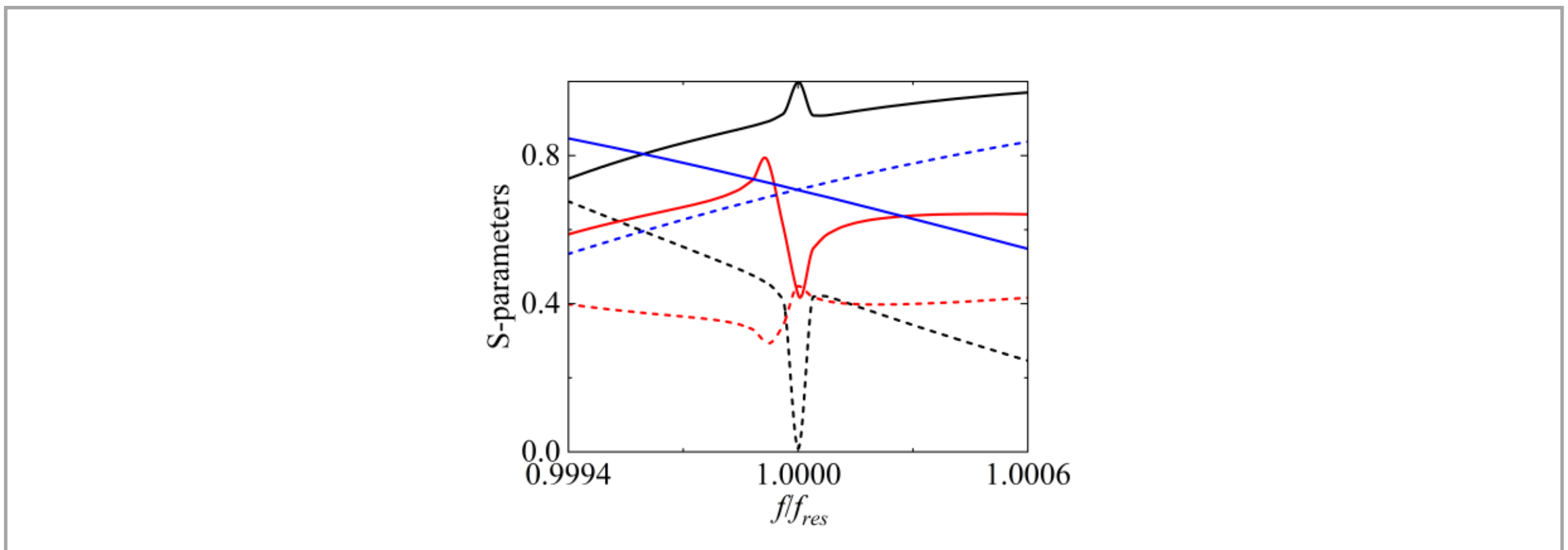

**Figure 8.** Polarization-dependent excitation at the electric intersection point ($\Delta h_d/h_d$ = -0.62). (a) Reflection and transmission responses for azimuth polarization angles $\varphi$ = (black line) 0°, (red line) 45° and (blue line) 90°, with resonant frequencies $f_{ATO,res}$ = 248.27 THz and $f_{TO,res}$ = 244.94 THz. The 0° polarization preferentially excites the electric ATO mode, while 90° favors the electric TO.

To illustrate all four resonances within a single spectral window, table 1 shows unnormalized responses at at $\Delta\, h_d/h_d$ = -0.3, an asymmetry value where all modes are simultaneously observable. The responses clearly resolve the magnetic ATO and TO, electric TO and ATO. The complementary polarization selectivity is evident: 0° excitation predominantly addresses the magnetic ATO and electric TO modes, while 90° favors the magnetic TO and electric ATO modes. All resonances are in Supplementary Material, section S8, table 10.

**Table 1.** Frequency and Q-factors of perturbed collective electric and magnetic toroidal eigenmodes.

| # | Mode type | Polarization | Frequency | Q-factor |
|---|---|---|---|---|
| 1 | H-ATO | 90° | 240.44 | $1.3\times10^4$ |
| 2 | H-TO | 0° | 242.8 | $1\times10^4$ |
| 3 | E-TO | 90° | 244.94 | $3.3\times10^4$ |
| 4 | E-ATO | 0° | 248.27 | $5.1\times10^4$ |

Overall, these results demonstrate efficient far-field access to high-Q toroidal resonances with strong field localization. The observed polarization-dependent coupling provides a simple external tuning mechanism (requiring no structural modification) that is advantageous for dynamically reconfigurable or multiplexed metasurface devices [19,21–27,32].

For comparison with studies on metasurfaces composed of a single nanorod, the present system operates in a mode where both field enhancement and Q-factors are simultaneously high. This performance surpasses emission-control metasurfaces ($Q$ = 270 in [1] and $Q$ = 1500 in [2]) as well as electro-optic tunable designs (electric field of 0÷300 V/m and magnetic field of 0÷4 A/m in [37]; electric field of 0-5×$10^6$ V/m in [38]; electric field of 3×$10^7$ V/m in [39]). Such improvements enhance sensitivity of the structure to variations in material parameters.

**3.4. Robustness of toroidal quasi-BICs and material losses impact**

Having established polarization-controlled excitation of toroidal resonances, we next assess their robustness to realistic material losses (a crucial aspect for practical applications).

Figure 9 presents the normalized responses for the $C_s^{(1)}$ configuration at two $\Delta h_d/h_d$ and $\varphi$, with Si loss tangents of 0, $10^{-4}$, and $10^{-3}$, revealing the differential impact of material losses.

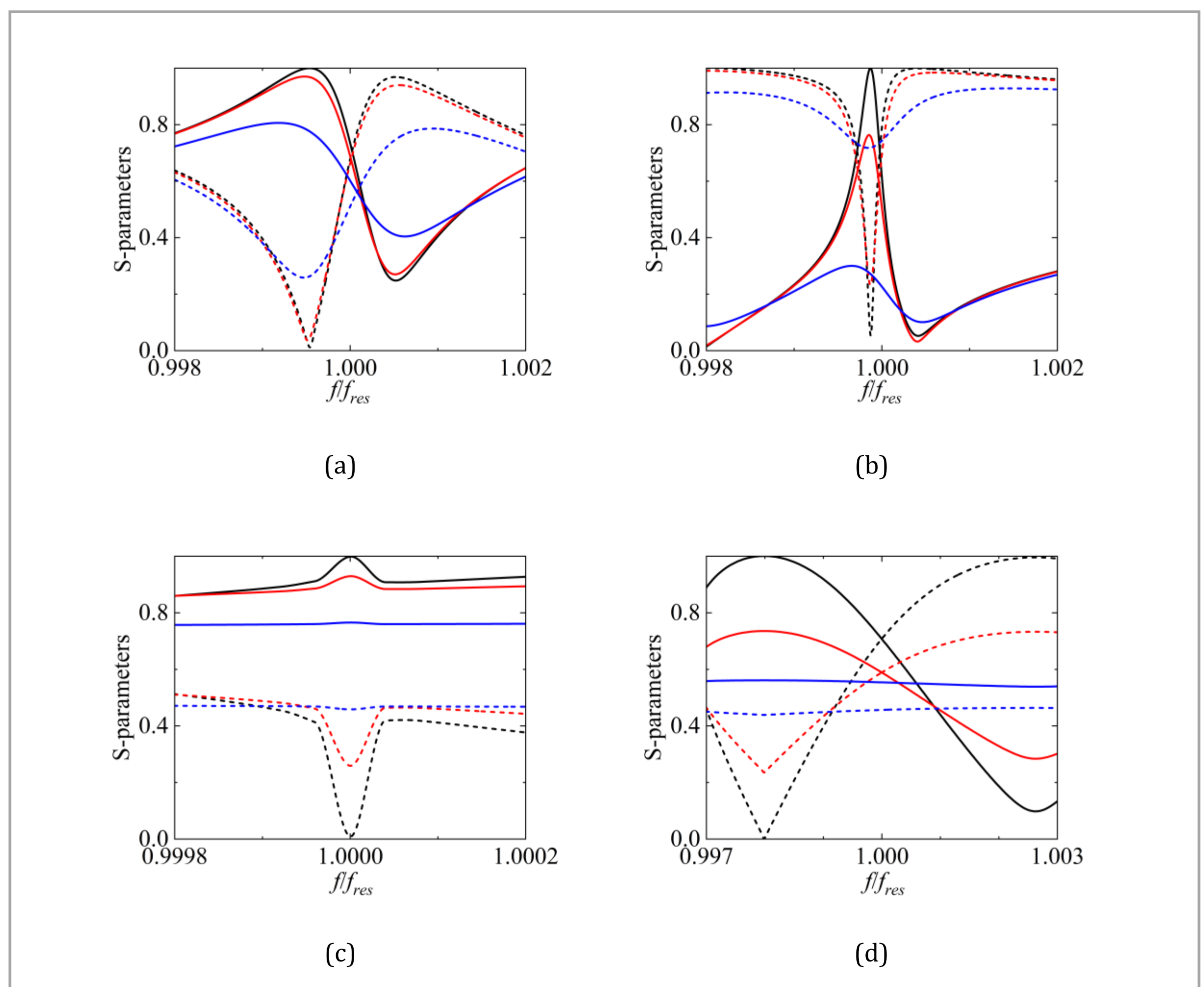

**Figure 9.** Normalized reflection (solid) and transmission (dashed) responses for varying Si loss tangent (black line: 0, red line: $10^{-4}$, blue line: $10^{-3}$). (a) $\Delta h_d/h_d$ = -0.44, $\varphi$ = 0° (magnetic TO). (b) $\Delta h_d/h_d$ = -0.44, $\varphi$ = 90° (magnetic ATO). (c) $\Delta h_d/h_d$ = -0.62, $\varphi$ = 0° (electric ATO). (d) $\Delta h_d/h_d$ = -0.62, $\varphi$ = 90° (electric TO).

A consistent hierarchy of robustness to material losses emerges from figure 9. The magnetic TO mode, in figure 9(a), exhibits the greatest resilience, maintaining well-defined resonances even at $10^{-3}$ loss tangent, with only moderate amplitude reduction and linewidth broadening. The magnetic ATO mode, in figure 9(b), shows the next level of robustness, with more pronounced degradation than the magnetic TO but still retaining clear resonant features at $10^{-3}$ loss. The electric ATO mode, in figure 9(c), displays intermediate sensitivity, with significant degradation at $10^{-3}$ loss but a discernible resonance remaining. In contrast, the electric TO mode, in figure 9(d), is the most loss-sensitive, with near-complete disappearance of the resonance at $10^{-3}$ loss tangent. Thus, the hierarchy of loss sensitivity from greatest resilience to most sensitive is: magnetic TO > magnetic ATO > electric ATO > electric TO. This hierarchy correlates with the spatial energy distribution of each mode. TO modes concentrate electromagnetic energy within dielectric volumes, where dissipative losses are inherently lower. ATO modes, by contrast, exhibit enhanced field localization at the interfaces between adjacent trimers, where losses are more pronounced due to stronger field gradients and proximity to air gaps. The electric TO mode (the most loss-sensitive) shows the strongest interfacial field concentration, while the magnetic TO mode (the most robust) confines energy predominantly inside the silicon nanorods. The magnetic ATO and electric ATO modes occupy intermediate positions, consistent with their mixed energy distribution between dielectric volumes and interfaces. For all configurations, increasing the loss tangent from 0 to $10^{-3}$ progressively reduces resonance amplitude and broadens linewidths, while resonance frequencies remain stable. This confirms that spectral positions are determined by geometry and collective modal structure rather than dissipative effects. These results establish clear design guidelines for practical implementations. For applications requiring tolerance to realistic material losses, operation with magnetic TO mode excitation ($\Delta h_d/h_d$ = -0.44, $\varphi$ = 0°) is strongly recommended. If electric mode operation is required for specific functionalities, the electric ATO mode ($\Delta h_d/h_d$ = -0.62, $\varphi$ = 0°) offers superior loss tolerance compared to its electric TO counterpart. These trade-offs must be carefully considered when designing high-Q dielectric metasurfaces for real-world applications with non-negligible material absorption.

### 3.5. Influence of substrate and protective layers on toroidal intersections

While the preceding analysis considered a free-standing metasurface, practical devices often require substrates or protective coatings. We therefore examine how these additional layers affect the TO/ATO intersections. Introducing a substrate modifies the spectral landscape. Two additional cases with a silicon dioxide substrate (one with silicon nanorods and another with matched silicon dioxide nanorods) exhibited no intersection of the TO and ATO modes.

For both silicon and silicon dioxide substrates, the TO and ATO eigenfrequencies shift and remain spectrally separated, indicating that a low-index environment inherently lifts the degeneracy and suppresses mode intersections. These observations establish symmetry breaking as a necessary but not sufficient condition for TO/ATO crossings. To maintain compatibility with realistic fabrication while preserving the intersection behavior, we considered a high-index silicon metasurface ($\varepsilon$ = 12) covered with protective layers of different materials. Table 2 summarizes the simulation results for the magnetic TO–ATO intersection point ($\Delta h_d/h_d$ = -0.44) with and without protective coatings.

Perfluorodecyltrichlorosilane (FDTS) [$CF_3(CF_2)_7(CH_2)_2SiCl_3$] and Hexamethyldisilazane (HMDS) [$Si_2N(CH_3)_6$] were selected as representative protective layers due to their common use in dielectric metasurface fabrication. Both layers are ultrathin (2 nm) with low refractive indices (2.2–2.3) and small imaginary permittivity (0.001), ensuring minimal perturbation of the electromagnetic field distribution.

From the results in table 2, several conclusions can be drawn. First, both FDTS and HMDS preserve the TO–ATO intersection condition—the degeneracy remains intact despite the presence of the protective layer. Second, the resonance frequency shifts slightly downward relative to the bare case (from 243.76 THz to 242.87 THz for FDTS and 242.8 THz for HMDS), consistent with the increased effective permittivity of the surrounding medium. Third, the Q-factors remain unchanged at $4\times10^3$ for the degenerate modes, indicating that the low-loss protective coatings do not introduce significant additional dissipation.

**Table 2.** Simulation results for the magnetic TO–ATO intersection point ($\Delta h_d/h_d$ = -0.44) with different protective layers.

| Si cover | None (air) | FDTS | HMDS |
|---|---|---|---|
| $h_{cover}$, nm | 0 | 2 | 2 |
| $Re(\varepsilon)$ | 1 | 2.2 | 2.3 |
| $Im(\varepsilon)$ | 0 | 0.001 | 0.001 |
| $f_{ATO}$ = $f_{TO}$, THz | 243.76 | 242.87 | 242.8 |
| $Q_{ATO}$ | $4\times10^3$ | $4\times10^3$ | $4\times10^3$ |
| $Q_{ATO}$ | $10^3$ | $10^3$ | $10^3$ |

Among the two materials, FDTS provides the most effective protective layer. Its resonance frequency of 242.87 THz lies closer to the bare-silicon value than HMDS (242.8 THz), and its asymmetry parameter $\Delta h_d/h_d$ = -0.44 exactly matches the unprotected case. This finding confirms that the use of an FDTS layer justifies adopting a simplified bare-silicon model in subsequent studies of symmetry-breaking and spectral characteristics, while still preserving consistency with realistic protective-layer conditions.

## 4. Discussion

The systematic comparative analysis presented in this work establishes a unified framework for understanding electric and magnetic toroidal resonances [4,5,31–36] in all-dielectric hexagonal trimer metasurfaces [18,19,21–27]. Our results reveal both profound similarities and essential differences between these two mode families, with important implications for the design of high-Q photonic devices. Compared to previous studies focusing primarily on magnetic toroidal modes, the present work provides a unified analysis of both electric and magnetic toroidal excitations within the same metasurface geometry, enabling direct comparison of their spectral, polarization, and loss characteristics.

The one-to-one correspondence between electric and magnetic toroidal mode pairs (figures 2-3) establishes a predictive design principle rooted in electromagnetic duality. Magnetic modes offer higher Q-factors but greater asymmetry sensitivity, while electric modes provide better Q-factor stability under moderate symmetry breaking [1,32,38]. Polarization-selective excitation (figures 7–9) enables dynamic switching between mode families, with reversed ordering for electric modes relative to magnetic ones [31,35–38]. The loss hierarchy correlates with spatial energy distribution, revealing that dual counterparts exhibit opposite robustness.

For practical implementations, magnetic TO offers the most robust operation, while the electric ATO provides the best loss tolerance when electric field dominance is required.

These distinct near-field topologies translate into experimentally accessible signatures. Specifically, polarization-resolved far-field spectroscopy can selectively excite each mode (figures 7-9), while near-field scanning optical microscopy (NSOM) could directly map the field predicted in figures 2-3 [22]. This testability bridges our theoretical predictions with future experimental validation and provides clear guidelines for distinguishing electric from magnetic toroidal responses in laboratory settings.

The dual electric-magnetic toroidal platform offers concrete advantages for several application areas [3,12,31]. The coexistence of two high-Q resonances (electric and magnetic) enables self-referenced detection: shifts in the electric mode report on local permittivity changes, while the magnetic mode provides a stable reference, suppressing thermal drift and environmental noise [31]. The Q-factors demonstrated here translate into figure-of-merit values for refractive index sensing, surpassing conventional metasurface sensors by several orders of magnitude.

For nonlinear optics [3,12], the spectral overlap of electric and magnetic resonances can enhance both second- and third-order nonlinear processes simultaneously. The electric toroidal modes concentrate fields in inter-nanorod gaps, ideal for nonlinear interactions with gain media or two-dimensional materials, while magnetic toroidal modes provide complementary field confinement within dielectric volumes. The ability to tune the frequency separation between electric and magnetic resonances via height modulation (figure 4) enables phase-matching conditions for sum-frequency generation and third-harmonic generation.

The orthogonal polarization selectivity demonstrated in figures 7-9 allows polarization-switchable routing of resonant energy. By operating at an intermediate asymmetry parameter (e.g., $\Delta h_d/h_d$ = -0.3 in table 1), both mode families are accessible with comparable Q-factors, enabling dynamic reconfiguration of the metasurface response simply by rotating the incident polarization—without any structural modification [12].

The height-modulation approach employed throughout this study is fully compatible with standard nanofabrication techniques, including multi-step electron beam lithography and controlled dry etching. The continuous tunability of $\Delta h_d/h_d$ through nanorod height adjustment offers practical advantages over lateral displacement schemes, as vertical geometry perturbations can be implemented with high precision and reproducibility across large-area metasurfaces. Furthermore, as demonstrated in Section 7, the use of an FDTS protective layer preserves the intersection behavior while providing compatibility with realistic fabrication conditions, justifying the simplified bare-silicon model used in our loss analysis. The design principles established here are scalable across the spectrum: by appropriate scaling of structural dimensions and material selection, the same platform can be translated to terahertz, mid-infrared, or visible wavelengths.

Our findings establish a general framework for symmetry-controlled modal interactions. Future work should explore oblique incidence [2,15,16], substrate effects, experimental validation, and simultaneous electric-magnetic toroidal excitation.

## 5. Conclusions

We have established a duality between electric and magnetic toroidal excitations in all-dielectric hexagonal trimer metasurfaces under controlled height-induced symmetry breaking. Using finite-element simulations, we identified and classified both electric and magnetic toroidal mode families within the same hexagonal supercell geometry, demonstrating that for every magnetic toroidal configuration there exists an electric counterpart with analogous symmetry properties and current topologies. The electric field of magnetic modes systematically maps onto the magnetic field of electric modes, establishing a duality that enables predictive design across mode families.

Key achievements include: (i) identification and classification of electric toroidal modes previously unexplored in the literature; (ii) demonstration of direct frequency intersections between electric TO and ATO modes, alongside magnetic TO/ATO intersections, yielding high-Q quasi-bound states in the continuum with Q-factors exceeding $10^4$; (iii) polarization-selective excitation enabling dynamic switching between mode families through the azimuth angle of incident wave; (iv) systematic analysis of material losses revealing a consistent hierarchy of robustness, directly linking loss sensitivity to spatial energy distribution and providing actionable design guidelines for real-world devices; (v) validation that FDTS protective layers preserve intersection behavior, bridging the gap between idealized simulations and practical fabrication. Thus, these achievements lead to three fundamental *novelty* conclusions that distinguish our work from previous studies on toroidal metasurfaces [17–27].

*First*, while prior works have treated electric and magnetic toroidal responses as separate phenomena (or have focused exclusively on magnetic toroidal modes [19–27]) our results demonstrate that these two families are intrinsically linked by electromagnetic duality. This duality is not merely a formal symmetry but a physically observable property: the electric field distribution of magnetic modes systematically maps onto the magnetic field distribution of electric modes, and both families follow identical symmetry rules under height modulation. Specifically, the magnetic field of the magnetic ATO mode mirrors the electric field of the electric TO mode, while the electric field of the magnetic TO mode corresponds to the magnetic field of the electric ATO mode. These complementary near-field topologies (electric fields concentrated in inter-nanorod gaps versus magnetic fields forming closed in-plane loops) provide a direct visual fingerprint of duality. No prior study has established such a one-to-one correspondence across both mode families within a single metasurface platform.

*Second*, the demonstration of frequency intersections between electric TO and ATO modes alongside magnetic TO/ATO intersections reveals that quasi-BIC formation via accidental degeneracy is a general phenomenon applicable to both electric and magnetic toroidal families (with distinct polarization-dependent energy redistribution at specific azimuthal angles). Previous reports of toroidal quasi-BICs have been limited to magnetic responses [4,5,31–36]. Our work extends previous studies by considering by showing that electric toroidal modes can also support high-Q states at their intersection points. Moreover, the two intersections occur at distinct asymmetry parameters, enabling selective access to different quasi-BIC families within the same device (a capability not previously demonstrated).

*Third*, the hierarchy of dielectric loss sensitivity (magnetic TO > magnetic ATO > electric ATO > electric TO) reveals a previously unrecognized connection between mode topology and dissipative robustness. While it is known that TO and ATO modes exhibit different spatial energy distributions [19], our quantitative analysis shows that this distinction translates into a systematic ordering of material loss tolerance that persists across both electric and magnetic families. Crucially, the most robust mode (magnetic TO) and the least robust mode (electric TO) are dual counterparts, demonstrating that duality in field topology does not imply equivalence in practical device performance. This finding provides actionable design guidelines absent from previous toroidal metasurface studies, where loss analysis has typically been limited to single-mode families or idealized lossless scenarios [1,2,4].

Beyond cataloging new modes, this work establishes that electric and magnetic toroidal responses are two complementary manifestations of the same symmetry, that quasi-BIC formation via mode crossing is universal across both families, and that loss sensitivity follows a predictable hierarchy governed by spatial energy distribution. These conclusions provide a basis for using toroidal duality as a design framework for high-Q ADMs.

The loss hierarchy has immediate practical implications: for maximum robustness, magnetic TO excitation is strongly recommended; for applications requiring electric field dominance, the electric ATO mode offers the best loss tolerance. These findings transform the abstract duality concept into a practical engineering tool for high-Q dielectric metasurfaces.

This work provides a unified framework for toroidal photonics, paving the way for next-generation high-Q devices in sensing, nonlinear optics, and quantum photonics.

## Acknowledgements

The authors sincerely thank Dr. Vladimir R. Tuz (V. N. Karazin Kharkiv National University; Jilin University) for his insightful problem statement, inspiring discussions, and expert guidance throughout this research. His deep expertise and constructive feedback were instrumental in shaping the methodology and enhancing the quality of this work.

Zoya Eremenko acknowledges the funding from the European Union under the Marie Skłodowska-Curie grant agreement no. MSCA4Ukraine project number 1.4 - UKR - 1232611 - MSCA4Ukraine (IFW Dresden). This project has received funding through the MSCA4Ukraine project, which is funded by the European Union. Views and opinions expressed are however those of the author(s) only and do not necessarily reflect those of the European Union, the European Research Executive Agency or the MSCA4Ukraine Consortium. Neither the European Union nor the European Research Executive Agency, nor the MSCA4Ukraine Consortium as a whole nor any individual member institutions of the MSCA4Ukraine Consortium can be held responsible for them.

We are grateful for support by the National Academy of Sciences of Ukraine under project "Kentavr-7" and are thankful to the IFW Dresden for its hospitality.

## Conflict of interest

The authors have declared that no competing interests exist.

## Data access statement

All data supporting the findings of this study are available within the article and its Supplementary Material.

## Ethics statement

This study does not involve human participants, human data, or animals, and therefore ethical approval was not required.

## Author contributions

O.A.Breslavets conceived the study, performed simulations, and wrote the manuscript. Yu.M. Savin and Z.E. Eremenko provided substantial editorial input, structured the manuscript plan, and incorporated key information to enhance its quality and align it with the standards of a Q1 journal.